\documentclass[journal]{IEEEtran}
\usepackage[T1]{fontenc}
\usepackage{mathtools}
\usepackage{xurl}
\usepackage{amsmath}
\usepackage{amssymb}
\usepackage{amsthm}
\usepackage{amsfonts}
\usepackage{braket}

\usepackage{upgreek}
\usepackage{dirtytalk}

\usepackage{optidef}

\usepackage{algorithmic}
\usepackage{algorithm}
\usepackage{array}
\usepackage[caption=false,font=normalsize,labelfont=sf,textfont=sf]{subfig}
\usepackage{textcomp}
\usepackage{stfloats}
\usepackage{url}
\usepackage{verbatim}
\usepackage{graphicx}
\usepackage{setspace}
\usepackage{calc}
\usepackage{enumerate}
\usepackage[usenames]{color}

\usepackage{amssymb}
\usepackage{upref}
\usepackage{epic,eepic}
\usepackage{times}
\usepackage{dsfont}
\usepackage{comment}
\usepackage{dirtytalk}
\usepackage{braket}
\usepackage{longtable}
\usepackage{hyperref}

\usepackage{xcolor}
\usepackage{xpatch}

\ifCLASSINFOpdf
\else
\fi
\usepackage{cite}

\begin{document}
%

\title{Physics-Informed Quantum Communication Networks: A Vision Towards the Quantum Internet}

\author{\IEEEauthorblockN{Mahdi Chehimi and Walid Saad}
\IEEEauthorblockA{Wireless@VT, Bradley Department of Electrical and Computer Engineering, Virginia Tech, Arlington, VA USA,\\
Emails: \{mahdic,walids\}@vt.edu}}

\maketitle

\begin{abstract}
Quantum communications is a promising technology that will play a fundamental role in the design of future networks. In fact, significant efforts are being dedicated by both the quantum physics and the classical communications communities on developing new architectures, solutions, and practical implementations of quantum communication networks (QCNs). Although these efforts led to various advances in today's technologies, there still exists a non-trivial gap between the research efforts of the two communities on designing and optimizing the performance of QCNs. For instance, most prior works by the classical communications community ignore important quantum physics-based constraints when designing QCNs. For example, many existing works on entanglement distribution do not account for the decoherence of qubits inside quantum memories and, thus, their designs become impractical since they assume an infinite lifetime of quantum states. In this paper, we bring forth a novel analysis of the performance of QCNs in a physics-informed manner, by relying on the quantum physics principles that underly the different components of QCNs. The need of the physics-informed approach is then assessed and its fundamental role in designing practical QCNs is analyzed across various open research areas. Moreover, we identify novel physics-informed performance metrics and controls that enable QCNs to leverage the state-of-the-art advancements in quantum technologies to enhance their performance. Finally, we analyze multiple pressing challenges and open research directions in QCNs that must be treated using a physics-informed approach to lead practically viable results. Ultimately, this work attempts to bridge the gap between the classical communications and the quantum physics communities in the area of QCNs to foster the development of the future communication networks towards the quantum Internet.
\end{abstract}


%
\IEEEpeerreviewmaketitle

\section{Introduction}\label{sec_intro}
\IEEEPARstart{F}{uture} communication systems must handle massive volumes of data and sophisticated applications like digital twins which require significant advances in communications and computing technologies. However, classical computing capabilities have become saturated due to imminent end of Moore's law. This motivates re-engineering current computing technologies to ones that can serve future out-of-the-box computing-hungry applications. Here, quantum computing can enable tremendous speedups in executing complex algorithms and analyzing high-dimensional data. To incorporate quantum computers in communication systems, quantum communication networks (QCNs) must be developed side-by-side with classical networks. So, \emph{why do we need QCNs?} The answers to this question stem from key advantages:



\paragraph*{\textbf{Unbreakable Security}} One fundamental advantage of quantum communications is strong, nearly unbreakable security -- a feature necessary to support mission-critical applications. 

\paragraph*{\textbf{Higher Data Throughput}} A second distinguishing feature of QCNs is their promising potential for achieving increased throughput and information capacity. This is grounded on the concept of quantum superposition and higher order quantum states \cite{chowdhury2020_6g_quantum_1}. For example, quantum embeddings using quantum feature maps can be used to encode classical data into quantum states, which could enhance the transmission rate. 

\paragraph*{\textbf{Creating Computing-intensive Networks}}
Another unique QCNs' feature is their unprecedented computational capabilities. Quantum computing nodes can enable sophisticated artificial intelligence (AI) and network control algorithms. Hence, having QCNs facilitates shifting the current communication models from a communication-intensive structure to a computing-intensive one. Particularly, quantum technologies potentially enable extracting complex data patterns, which could be used for semantic communications.

These important features make QCNs a pillar of a much coveted, global \say{Quantum Internet} (QI) \cite{quantum_internet1} that could allow any two points on Earth to execute quantum communication services. However, designing QCNs faces multiple challenges ranging from the successful generation of high-quality quantum states, to overcoming their inherent decoherence.

Many prior works~\cite{kozlowski2019towards,gyongyosi2018multilayer_optimization,wei2022towards,quantum_internet1} that consider QCNs either focus on summarizing their components and challenges \cite{kozlowski2019towards,gyongyosi2018multilayer_optimization}, in isolation, or survey practical QCN implementations and enabling technologies \cite{wei2022towards}. Meanwhile, other works \cite{quantum_internet1} investigate QCN applications and their deployment challenges. However, most of these works ignore the challenge of making impractical physics-uninformed assumptions in QCN designs. In contrast to~\cite{kozlowski2019towards,gyongyosi2018multilayer_optimization,wei2022towards,quantum_internet1}, we propose a holistic physics-informed framework for the design, analysis, and optimization of QCNs, while identifying the underlying fundamental challenges. This is important to unite the efforts towards designing practical QCNs firmly grounded on quantum physics principles. The proposed framework (see Figure \ref{fig_contributions}) provides a building block towards aligning the perspectives of the quantum physics and the classical communications communities for practically developing hybrid quantum-classical networks and, ultimately, the QI with applications like quantum key distribution (QKD) and quantum sensing.



\subsection{Contributions}
\begin{figure}[t]
\begin{center}
\centerline{\includegraphics[width=\columnwidth]{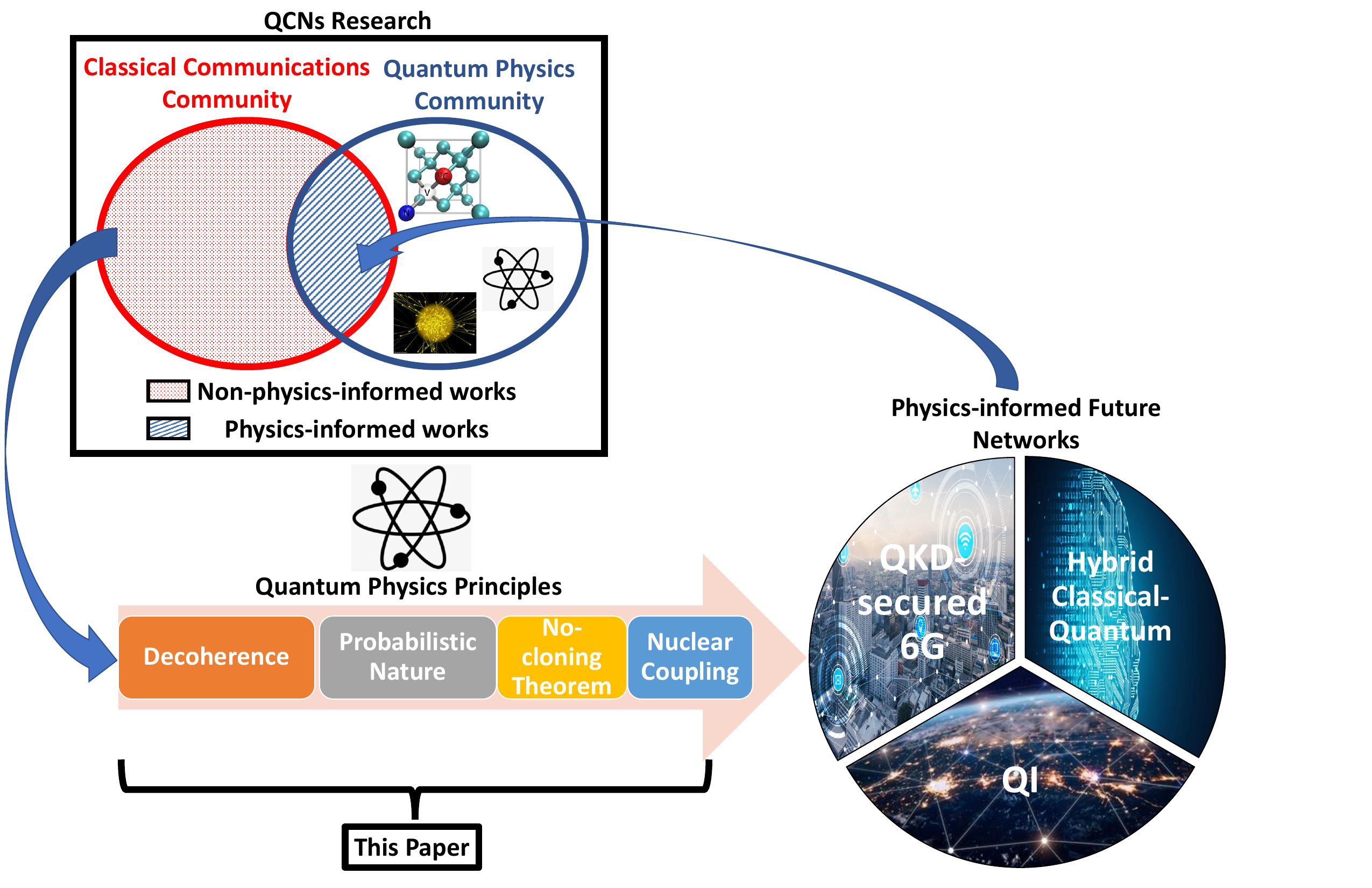}}
\caption{Evolution towards physics-informed QCNs}
\label{fig_contributions}
\end{center}
\vskip -0.4in
\end{figure}

Our goal is to bridge the gap between the different research directions for designing QCNs adopted by the classical communications and the quantum physics communities. Our contribution is therefore a novel framework, dubbed \emph{physics-informed QCNs}, for designing, analyzing, and optimizing the performance of QCNs based on their governing quantum physics principles. Towards this goal, we make key contributions (see Figure \ref{fig_contributions}):
\begin{itemize}
    
    
    \item We analyze multiple QCN research challenges that are currently being tackled while ignoring their governing quantum physics principles. Then, we discuss a physics-informed QCN analysis that could guarantee developing practical QCN designs. For each challenge, we identify the distinguishing features of the physics-informed framework compared to existing non-physics-informed approaches.
    
    \item We identify novel quantum physics-based performance metrics and controls to enhance QCN performance while adhering to the practical restrictions imposed by quantum physics.

    \item We explore the role of quantum information processing in generating high-quality quantum signals and extracting semantic features from classical data using physics-informed QCNs. In addition, we investigate the QCN resource management problem with a physics-informed approach using the identified quantum controls and metrics.

\end{itemize}

\section{Physics-informed QCNs: Motivation and Challenges}\label{sec_physics_informed}

QCNs will inevitably be integrated with the classical Internet to form a powerful classical-quantum Internet. Henceforth, future communication systems require simultaneous advances in the design and analysis of both QCNs and classical networks. 

However, different, unaligned approaches exist for handling QCNs. To explore their differences, we now analyze the current approaches for studying QCNs and discuss their drawbacks. Then, we introduce the concept of \emph{physics-informed QCNs}, explain why it is necessary, and analyze how it could overcome several QCN challenges.

\subsection{What does physics-informed mean? Why is it needed?}

Currently, there are two mainstream directions for studying QCNs. One focuses on the physical hardware layer, and practically deploying quantum mechanics in QCNs. This direction is mainly led by the quantum physics community. However, in this area, only a handful of works by this community \cite{pant2019routing} consider QCN networking scenarios where quantum resources are managed, and scheduled, and the QCN performance is optimized analogously to classical networks. Such analyses are necessary to develop the QI. 

In contrast, the second approach for tackling QCNs, led by the classical communications community, mainly focuses on analyzing diverse QCN networking problems while often being incognizant to the quantum hardware and its restrictions. While some exceptions exist (e.g., \cite{pant2019routing}), most works \cite{gyongyosi2018multilayer_optimization,schoute2016shortcuts,quantum_queuing_delay,rezai2021quantum,chen2019quantum,li2021building} from this community share the common drawback of \emph{not being guided by the quantum physics principles, and making impractical assumptions}. For example, in \cite{quantum_queuing_delay}, the authors ignore the principle of quantum decoherence in quantum memories, and the work in \cite{rezai2021quantum} ignores any physical QCN losses.

Consequently, there is a gap between the quantum physics and classical communications communities regarding QCN development. Particularly, the quantum physics community lacks the expertise on networking applications and performance optimization, while the classical communications community lacks the physics and hardware-related experience of the quantum physics community needed to ensure that their QCN designs consider real-world limitations. In fact, there is no comprehensive, holistic framework that lays down the foundations needed to develop QCN designs and networking analyses while relying on quantum physics. Such a \emph{physics-informed} framework for QCNs -- which we introduce -- must account for the physical limitations of the various QCN components and operations to ensure practicality.

\subsection{QCNs Challenges Requiring Physics-informed Approaches}\label{sec_challenges_requiring}
\begin{figure*}
\includegraphics[width=\textwidth]{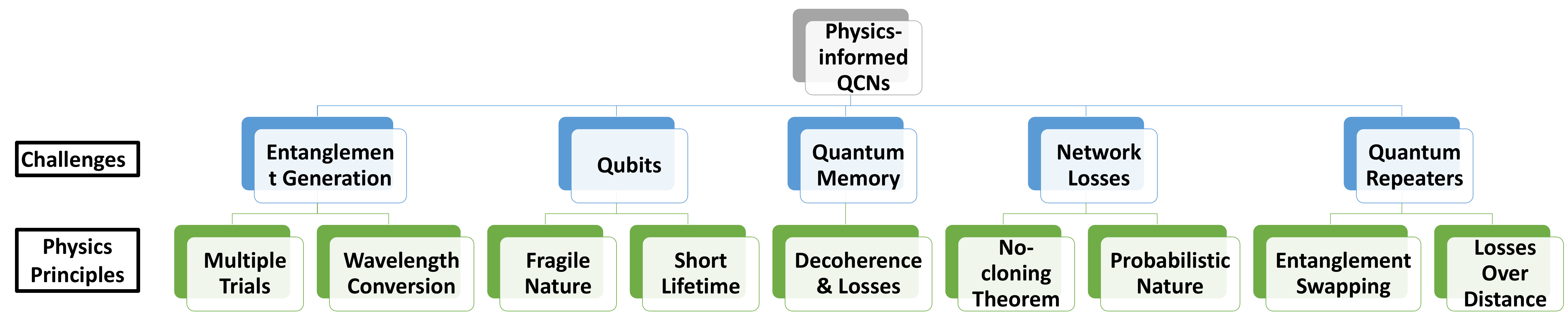}
\caption{Challenges requiring a physics-informed approach}
\label{fig_areas}
\vskip -0.2in
\end{figure*}

Next, we analyze the research challenges stemming from the aforementioned gap (see Figure \ref{fig_areas}), and we discuss how our physics-informed approach presents an opportunity to develop more practical QCNs.

\subsubsection{Entanglement Generation Process}\label{sec_entanglement_generation_challenge}
In classical communications, information bits are easily transferred as electromagnetic waves using oscillators at certain frequencies. Thus, the generation of information carriers does not affect the networks' performance. In contrast, in QCNs, the information carriers, qubits or quantum states, are physically realized using more fragile technologies like photons in which information is encoded. To transmit the information between quantum nodes, an \emph{entangled} pair of qubits is usually generated \cite{nielsen_book}. However, from a physics perspective, the generation of entangled qubits is nontrivial and has a probabilistic nature following the uncertainty principle that must be accounted for. Particularly, a single-photon source (SPS) performs multiple trials before successfully generating an entangled pair of qubits. Moreover, the generated qubits have a limited, short coherence lifetime. Further, an emitted photon may not fall into the telecommunications wavelength bands (e.g., fiber optical links), thus requiring frequency conversion, which is a lossy process \cite{kozlowski2019towards}.  

Many existing works consider the entanglement generation process while ignoring its quantum physics nature. One example is the assumption of perfect and instantaneous entanglement generation \cite{gyongyosi2018multilayer_optimization}. Such assumptions could lead to poor quality of the quantum states and inaccurate QCN delay expectations.

\subsubsection{Network Losses and Errors}
In classical communications, most error correction techniques use redundant codes. In contrast, in QCNs, this approach is not possible due to the quantum physics principles underlying QCNs. Examples include the \emph{No-cloning theorem} and the \emph{No-broadcast theorem}, which state that an arbitrary quantum state cannot be copied or broadcasted to multiple destination simultaneously \cite{nielsen_book}. Further, QCNs have a unique source of losses compared to classical networks, stemming from the principle of quantum decoherence, where quantum states collapse over time. 

Thus, a physics-informed QCN design must account for their inevitable losses. In fact, there are various examples of designs ignoring the physics-based losses in QCNs. This includes the assumption of a lossless QCN structure where pure states stay pure and a quantum state does not undergo any losses \cite{rezai2021quantum}, and the assumption of perfect quantum state transfer with fidelity 1 \cite{chen2019quantum}.

\subsubsection{Quantum Repeaters' Designs and Challenges}
Classical repeaters can amplify and forward the received signals to their next destination. In contrast, quantum repeaters cannot adopt this approach due to the No-cloning theorem. Generally, quantum repeaters use concepts like entanglement swapping, where a probabilistic Bell state measurement (BSM) operation is performed between two qubits, each of which is entangled with another qubit, to generate a new entanglement connection between two distant quantum nodes. Hence, the long communication distance is split into shorter segments with repeaters performing entanglement swapping, which helps overcoming photon losses that increase with the travelled distance. Moreover, since the two qubits used in entanglement swapping protocols may not arrive simultaneously, then quantum repeaters have to store and retrieve entangled qubits in quantum memory.

Evidently, the design and operation of quantum repeaters is more challenging than classical repeaters due to various losses and probabilistic operations. One example \cite{li2021building} that neglects those physics principles is the consideration that quantum repeaters are perfect devices with lossless, perfect quantum memories, which leads to poor QCN performance.

\subsubsection{Qubits lifetime}
In classical networks, information bits can be preserved forever in highly-reliable solid state memories, and can wait in long queues at access points with minimal losses. In contrast, in QCNs, quantum states are fragile and suffer from quantum decoherence due to their interaction with the surrounding environment, which could cause the quantum states to collapse and lose the data. Although advancing today's quantum technology could dramatically increase the coherence time of qubits, it will not be able to completely overcome the decoherence losses, particularly when qubits are stored in long queues in large-scale QCNs.

Thus, although the storage and preservation of quantum states challenges are generally a technology deficiency in today's QCNs, they also must remain cognizant of the underlying quantum physics. For instance, if the storage of qubits were assumed to be lossless with infinite qubits' lifetime \cite{schoute2016shortcuts}, then the resulting QCN design and performance will face serious challenges in terms of scheduling and resource management, along with impractical delay and throughput results. 

\subsubsection{Quantum Memory Capacity and Storage Time}\label{sec_II_memory}
In classical communications, effective storage of data bits is achievable using solid state memories with tremendous storage capabilities, and a nearly infinite storage time. In contrast, in QCNs, building and designing an efficient, lossless quantum memory is nontrivial and governed by unique quantum physics principles. Specifically, quantum states stored in quantum memories suffer from \emph{decoherence}, so they have a limited storage time. Moreover, quantum memories do not have a very large capacity, and cannot generally process and retrieve many quantum states simultaneously (this varies with adopted quantum technology).

Those challenges can be significantly reduced as the quantum technologies advance in the future. However, due to different probabilistic operations and measurements that are performed on qubits inside quantum memories, such qubits will still suffer from multiple losses due to their intrinsic quantum physics nature. Henceforth, it is important to adopt a physics-informed approach when considering quantum memories in QCNs, while also being aware of the technology capabilities. An example of ignoring the physics principles in quantum memories is the assumption of a perfect memory where quantum states are stored, retrieved, and transferred without experiencing any losses from their surroundings \cite{quantum_queuing_delay}. Such premises lead to poorly designed entanglement routing and scheduling protocols due to ignoring photon losses, decoherence, and the actual qubits' quality.

In a nutshell, from the aforementioned challenges, one can see the importance of studying QCNs through the lens of a physics-informed framework that aligns the visions of both the quantum physics and the classical communications communities.


\vspace{-0.15cm}
\section{Physics-informed Problems and Future Directions in QCNs}\label{sec_problems}
\begin{figure}[t]
\begin{center}
\centerline{\includegraphics[width=\columnwidth]{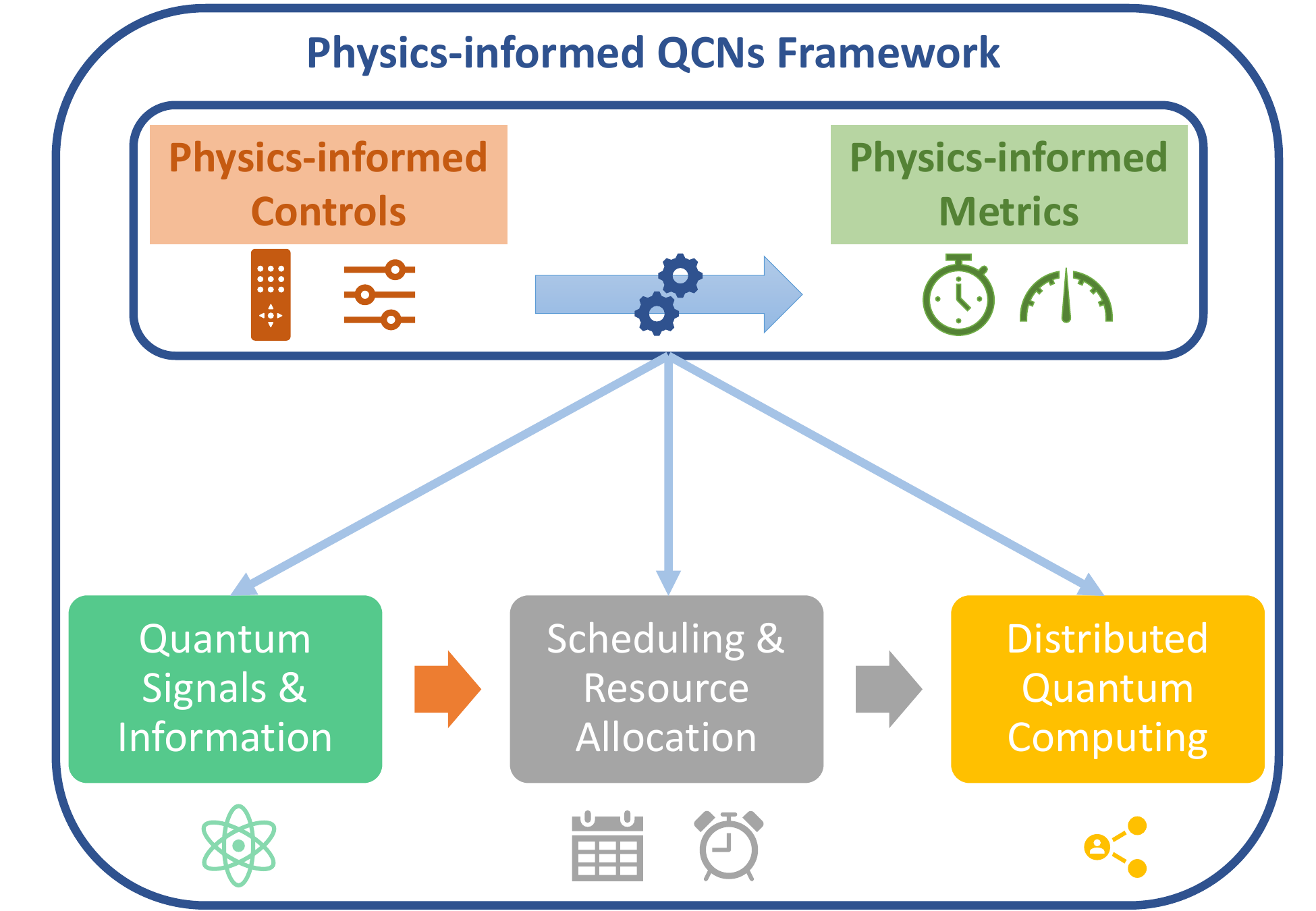}}
\caption{Physics-informed future research directions}
\label{fig_future_directions}
\end{center}
\vskip -0.35in
\end{figure}

\vspace{-0.1cm}
Now, we first identify various physics-informed performance metrics and controls necessary for the design and analysis of QCNs, then we use them in the preparation of quantum signals and information processing. Using those quantum signals, physics-informed scheduling and resource management algorithms are designed. We conclude by exploiting the proposed principles, signals, and scheduling algorithms in performing physics-informed distributed quantum computing, as in Figure \ref{fig_future_directions}.

\subsection{Physic-informed Performance Metrics}\label{sec_metrics}
In classical networks, metrics such as rate and delay quantify performance. Meanwhile, wireless channel impediments like fading and interference can hinder classical networks' performance. To develop a physics-informed framework for QCNs, we identify and analyze the analogous metrics and impediments in QCNs by relying on the quantum physics principles. We also explain why those metrics have a unique tangible physical meaning.





\subsubsection{Throughput}\label{sec_metric_throughput}
After the successful generation of entangled pairs of qubits at a certain entanglement generation rate, the achieved throughput by a QCN will correspond to the number of entangled qubits successfully preserved and delivered to each user. Users may have heterogeneous requirements, thus requiring a different amount of entangled qubits depending on the application \cite{chehimi2021entanglement}. Moreover, the achieved throughput and the entanglement generation rate (i.e., the rate at which an SPS tries to generate entangled qubits) are closely related to the adopted quantum technology. As an illustrative example, each technology takes a different time duration between every two entanglement generation attempts, which directly affects the QCN performance.

\subsubsection{Fidelity} \label{sec_fidelity}
When a qubit arrives at a receiving node, it should ideally arrive in its original state. However, due to the different compression-decompression procedures, and the various losses experienced during transmission, a distorted version of the quantum state may be received. \emph{Quantum fidelity} quantifies this distortion and captures how close a quantum state is to another state. Using this measure, the receiver can identify how close the received state is to its expected original state. 

\subsubsection{Quality of Matter Qubits (QoMQ)}\label{sec_metric_QoMQ}
Matter qubits, which represent a practical realization of a quantum memory, store quantum states related to different QCN nodes. The quality of the available matter qubits significantly affects the QCN performance, and it can be enhanced by optimizing network parameters. The QoMQ may be measured by three factors:
\begin{itemize}
    \item Efficiency of the absorption of quantum states into the matter qubit.
    \item Coherence time, which is the time duration during which the matter qubit can maintain a stored quantum state without losing it.
    \item Efficiency of extraction of quantum states back from the matter qubit.
\end{itemize}
The physical realizations of matter qubits include a wide range of technologies, such as nitrogen-vacancy (NV) centers (particularly the nuclear spins), trapped ions, and rare earth materials. Each such technology incorporates a compromise between the different characteristics specifying the quality of the quantum memory \cite{kozlowski2019towards}.   



\subsubsection{Delay}\label{sec_metric_delay}
Another fundamental QCN measure of performance is the delay associated with the operations and steps needed for quantum communications. As mentioned in Section \ref{sec_II_memory}, a quantum memory has a limited lifetime, which introduces major challenges on controlling QCNs. There exist multiple sources of delay in QCNs: 1) Delay from waiting until a successful entangled photon is emitted from the SPS, 2) Delay caused by waiting until a graph quantum state is successfully generated, in case quantum error correction (QEC) techniques are used (see Section \ref{sec_logical_states}), 3) Delay from waiting until the BSM outcome is heralded back to the transmitter, 4) Delay from repeating the quantum purification process until achieving a certain high fidelity-level \cite{nielsen_book}, and finally, 5) Delays dependent on the quantum channel type, since, for example, communicating in free-space is faster than over fiber optics. Thus, every QCN operation must be accounted for in the delay analysis.

\subsubsection{Probability of Success (PoS)}\label{sec_PoS}
This is a general metric that applies at different QCN components, based on context. For instance, to quantify channel losses in an end-to-end quantum link, the PoS in the transmission of photons over quantum channels is a metric of particular significance. 

For some quantum operations, defining the PoS as a QCN variable parameter is challenging because of a dependence on certain uncontrollable hardware-related factors. For instance, the BSM operation of single photons has an almost $50\%$ PoS, which is totally caused by the measurement linear optics, and requires hardware-based solutions to enhance it. Further, the PoS for generating an entangled pair of qubits heavily relies on the SPs's hardware platform, as per Section \ref{sec_physics_informed}. Thus, when the PoS is considered, careful attention is needed to identify which QCN controls can enhance it.


\subsection{Physics-informed Controls in QCNs}\label{sec_controls}

\begin{figure}[t]
\begin{center}
\centerline{\includegraphics[width=0.8\columnwidth]{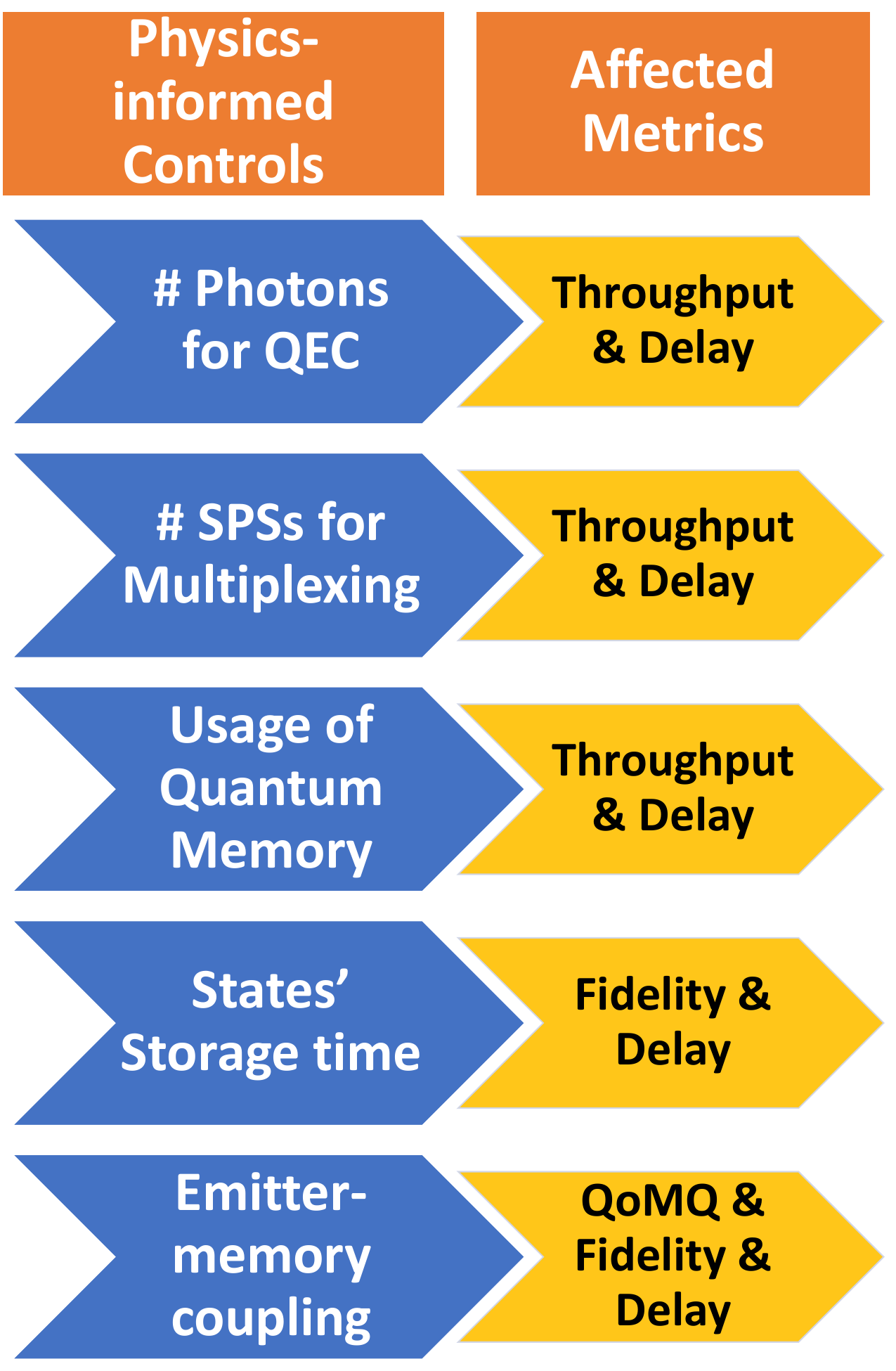}}
\vspace{-0.1in}
\caption{Physics-informed controls in QCNs} 
\label{fig_controls}
\end{center}
\vspace{-1.25cm}
\end{figure}

Classical networks include different control parameters for optimizing performance. Examples include wireless resources such as time, bandwidth, and energy. Analogously, QCN physics gives rise to many controllable parameters stemming from the application of QEC techniques, and the utilization of quantum memories. Meanwhile, some QCN components completely depend on the used hardware which restricts their performance by the technology's limitations. The main controllable resources in a QCN are summarized in Figure \ref{fig_controls}.

\subsubsection{Number of Photons in Logically-encoded Quantum States}\label{sec_logical_states}
QEC techniques are used to overcome the photon transmission losses over quantum channels. For instance, tree graph states can correct for photon losses, forming what is known as the \emph{logical encoding} of quantum states. In such states, the number of photons included in the quantum state is a controllable parameter that can be varied to optimize the QCN throughput and delay performance. From a physics perspective, each successful SPS photon emission is a probabilistic operation that takes a specific amount of time and trials until it succeeds, as discussed in Section \ref{sec_entanglement_generation_challenge}. Changing the number of photons in a logically-encoded quantum state changes the waiting time until the communication is performed, which directly affects the QCN delay. Moreover, since this number of photons is related to correcting errors in transmissions, then it directly impacts the throughput. 


\subsubsection{Number of SPSs for Multiplexing}\label{sec_multiplexing}
A promising QCN principle to enhance the PoS of establishing quantum links is \emph{multiplexing}, which is the counterpart of multiple access in classical networks. In this principle, multiple SPSs are used simultaneously to have better chances of successfully emitting photons. A quantum node, e.g., access point, coordinating a QCN must control the number of SPSs dedicated for each user based on its requirements. This directly affects the throughput of each user, and accordingly, the overall experienced delay prior to service. 

\subsubsection{Usage of Quantum Memory}\label{sec_control_memory_usage}
As per Section \ref{sec_II_memory}, quantum memory has a limited capacity, which must be carefully exploited by a quantum node to manage the stored quantum states. Particularly, quantum nodes must allocate a certain number of storage slots for the quantum states of each user. This management of the quantum memory significantly affects the achieved throughput \cite{quantum_queuing_delay} and the delay during service. 


\subsubsection{Storage Time of Quantum States}\label{sec_control_storage_time_memory}
As quantum states are stored for a longer duration, they start to decohere, which affects their quality and eventually results in losing the stored quantum information. A quantum node can control how long to store each qubit as well as which qubits must be prioritized based on their users' requirements. This time-scheduling of the quantum memory must be physics-informed by accounting for the decoherence and the quantum memory technology. This storage time of qubits in quantum memories is an important controllable parameter in QCNs since it directly affects the qubits fidelity, and the overall QCN delay.



\subsubsection{Emitter-Matter Qubit Coupling}\label{sec_coupling}

Each QCN node has an emitter (i.e., an SPS), that emits photons, and matter qubits that store quantum states, not necessarily photons. A diamond's NV center \cite{kozlowski2019towards} is a popular technology used in QCNs to realize emitters. In NV centers, the nuclear spins inside the diamond are the matter qubits representing the quantum memory. When a nuclear spin is used to store a quantum state, the remaining nuclear spins are dynamically decoupled from the electron spin to increase the lifetime of the chosen nuclear spin. Thus, only the desired matter qubit remains coupled to the electron spin of the NV center. 

Here, an interesting, physics-informed controllable parameter is the choice of the nuclear spin, the coupling of which is preserved with the electron spin of the NV center when dynamic decoupling is performed. This coupling directly affects the QoMQ since it changes the storage quality of the quantum memory. Moreover, it affects the QCN delay and fidelity. For instance, having a higher coupling (by choosing a closer nuclear spin to the electron spin) yields faster two-qubit gates needed for qubits' storage and retrieval, thus reducing delay. However, it also yields stronger unwanted interactions causing more losses, which affects fidelity. Meanwhile, when we have a weaker coupling, we reduce these losses, resulting in higher fidelity, at the expense of longer time delays to perform the two-qubit gates.




\subsection{Physics-informed Quantum Signals and Information Processing}
After identifying physics-informed metrics and controls, quantum signals are generated at the transmitter using photonic optical technologies. These signals, embedding quantum information, are then transferred to the receiver over quantum channels, and the receiver uses optical technologies to detect and identify them. Quantum optical technologies on both transmitter and receiver sides perform signal processing and computational tasks, which adds intelligence to QCNs. Such information processing requires a physics-informed approach since it should leverage the available QCN controls from Section \ref{sec_controls} to ensure the generation of high-quality quantum signals. For instance, quantum signals usually embed the information in the polarization of photons. Moreover, higher dimensional techniques, e.g., orbital angular momentum, may be employed to realize higher dimensional quantum states. In fact, the signal design directly affects the rate and the overall QCN communication quality. In this regard, metrics such as quantum fidelity and QoMQ play a fundamental role in quantifying the generated signals' quality.

Moreover, contextual communications is necessary for future applications, where the data semantics (i.e., meanings) are sent in lieu of large volumes of raw data. Having such semantic-aware communications is necessary to develop future AI-centric networks. Here, QCNs may play a transformative role in performing semantic communications more efficiently and accurately than their classical counterpart, by leveraging the unique physics-informed concepts of entanglement and fidelity. An important application of quantum information processing here pertains to embedding raw data in high-dimensional quantum states and undercovering hidden data patterns that represent their semantics, (see \cite{chehimi2022semantic} for preliminary results in this area). This task may be achieved by utilizing quantum clustering to discover special correlations between data samples and quantum state geometry to develop a common language or a vocabulary codebook for quantum-based semantics and goal-oriented communications. 


\subsection{Physics-informed Scheduling and Resource Management}\label{sec_scheduling}
Once the quantum signals are designed, the next step is to perform resource management and scheduling to optimize their usage at scale, since they are the main QCN resources. However, generating physics-uninformed, low-quality quantum signals would render the scheduling algorithms unable to satisfy the QCN users’ requirements. Henceforth, quantum schedulers must leverage different physics-informed control parameters, like entangled photons, quantum memory, and multiplexing SPSs (see Section \ref{sec_controls}) to ensure a fair resources' distribution across QCN users. An illustrative example is a quantum base station (QBS) serving multiple QCN users. This QBS has to tune its controls to optimize an objective function based on the aforementioned physics-informed performance metrics, e.g., to minimize delay or maximize throughput. The QBS must adopt a physics-informed approach in allocating a certain number of SPSs and dedicating specific stored entangled qubits to each QCN user so as to satisfy their requirements while considering the losses and challenges affecting the different quantum resources due to the physics principles. A mathematical exposition of such an example can be found in \cite{chehimi2021entanglement} where key simulation results are provided.

Most prior works (e.g., \cite{quantum_queuing_delay} and \cite{rezai2021quantum}) on QCN scheduling ignore multiple physics principles and rely on simplified, impractical assumptions such as perfect, infinite qubits storage in lossless quantum memories \cite{quantum_queuing_delay}, and lossless QCN architectures \cite{rezai2021quantum}. In contrast, scheduling and resource management in QCNs require a physics-informed approach due to unique quantum challenges. For instance, the task of scheduling QCN users and operations requires reliably storing the available resources for a long time. However, when considering a real QCN, this process is restricted by the short lifetime of quantum states. One solution is to modify state-of-the-art scheduling algorithms to incorporate the lifetimes of quantum states as constraints. The probabilistic nature of the entanglement generation process and other quantum operations such as BSM makes the scheduling problem in QCNs more sophisticated than its classical counterpart. Particularly, a quantum scheduler must balance between the storage time of quantum states and their fidelity, and must account for lost resources or failed quantum operations. One solution here is to apply (quantum) reinforcement learning algorithms to train the scheduler in a simulation environment before deploying it in practice. 
\vspace{-0.1cm}




\subsection{Physics-informed Distributed Quantum Computing over QCNs}
Future QI applications like distributed quantum computing and AI require distributing complex computations over networked quantum devices \cite{chehimi2021quantum}. Those applications require high-quality quantum signals that are managed in a physics-informed manner. In practice, distributed quantum computing is a key enabler for quantum cloud services, which have already been established in the industry. 

Technically speaking, relying on QCNs to perform distributed quantum computing is restricted by multiple physics-based constraints. For instance, synchronization between different QCN quantum devices is a critical challenge that affects the computing performance. This synchronization is closely related to the delay and the memory storage time controls. Thus, quantum scheduling algorithms depend on this synchronization and must consider it when dedicating memory storage slots for qubits. Moreover, the quality of quantum states is fundamental for distributed quantum computing applications. Particularly, distributing low-quality, distorted quantum states over QCNs is insufficient to consider the quantum communication successful, since such states may not be usable for the computing tasks. However, due to QCNs' underlying quantum physics, the task of maintaining a high quality of quantum states during transmission and storage becomes very challenging. Thus, the quantum fidelity must be carefully optimized in the physics-informed resource management protocols (see Section \ref{sec_scheduling}) to ensure effective distributed computing performance. 

Moreover, most quantum computing devices employ technologies such as quantum annealers, trapped ions, and superconducting technologies. In contrast, QCNs mainly use quantum photonic technologies whose underlying physics principles must be exploited to develop novel interfaces for connecting them. Such interfaces are necessary for distributed computing over QCNs since, for example, a quantum annealer's qubit cannot be distributed directly using QCN photons.

\vspace{-0.1cm}
\section{Conclusion}\label{sec_conclusion}
In this paper, we have performed a novel, holistic physics-informed analysis of QCNs that can help overcome various challenges hindering their deployment towards the QI. Particularly, the physics-informed approach bridges a significant gap that exists between the works of the quantum physics and the classical communications research communities on QCNs' designs and performance optimization in terms of compliance with quantum physics principles. Furthermore, we have identified physics-based performance metrics and control parameters necessary for enhancing the performance of QCNs. Finally, we have analyzed challenges and open QCN research directions that must be approached in a physics-informed manner.

\vspace{-0.1cm}


\bibliographystyle{IEEEtran}
\bibliography{References}

\begin{thebibliography}{10}
\providecommand{\url}[1]{#1}
\csname url@samestyle\endcsname
\providecommand{\newblock}{\relax}
\providecommand{\bibinfo}[2]{#2}
\providecommand{\BIBentrySTDinterwordspacing}{\spaceskip=0pt\relax}
\providecommand{\BIBentryALTinterwordstretchfactor}{4}
\providecommand{\BIBentryALTinterwordspacing}{\spaceskip=\fontdimen2\font plus
\BIBentryALTinterwordstretchfactor\fontdimen3\font minus
  \fontdimen4\font\relax}
\providecommand{\BIBforeignlanguage}[2]{{%
\expandafter\ifx\csname l@#1\endcsname\relax
\typeout{** WARNING: IEEEtran.bst: No hyphenation pattern has been}%
\typeout{** loaded for the language `#1'. Using the pattern for}%
\typeout{** the default language instead.}%
\else
\language=\csname l@#1\endcsname
\fi
#2}}
\providecommand{\BIBdecl}{\relax}
\BIBdecl

\bibitem{chowdhury2020_6g_quantum_1}
M.~Z. Chowdhury, M.~Shahjalal, S.~Ahmed, and Y.~M. Jang, ``6g wireless
  communication systems: Applications, requirements, technologies, challenges,
  and research directions,'' \emph{IEEE Open Journal of the Communications
  Society}, vol.~1, pp. 957--975, July 2020.

\bibitem{quantum_internet1}
M.~Caleffi, A.~S. Cacciapuoti, and G.~Bianchi, ``Quantum internet: From
  communication to distributed computing!'' in \emph{Proceedings of the 5th ACM
  International Conference on Nanoscale Computing and Communication}, Reykjavik
  Iceland, Sept. 2018, pp. 1--4.

\bibitem{kozlowski2019towards}
W.~Kozlowski and S.~Wehner, ``Towards large-scale quantum networks,'' in
  \emph{Proceedings of the Sixth Annual ACM International Conference on
  Nanoscale Computing and Communication}, Dublin Ireland, Sept. 2019, pp. 1--7.

\bibitem{gyongyosi2018multilayer_optimization}
L.~Gyongyosi and S.~Imre, ``Multilayer optimization for the quantum internet,''
  \emph{Scientific Reports}, vol.~8, no.~1, pp. 1--15, Aug. 2018.

\bibitem{wei2022towards}
S.-H. Wei, B.~Jing, X.-Y. Zhang, J.-Y. Liao, C.-Z. Yuan, B.-Y. Fan, C.~Lyu,
  D.-L. Zhou, Y.~Wang, G.-W. Deng \emph{et~al.}, ``Towards real-world quantum
  networks: A review,'' \emph{Laser \& Photonics Reviews}, vol.~16, no.~3, p.
  2100219, 2022.

\bibitem{pant2019routing}
M.~Pant, H.~Krovi, D.~Towsley, L.~Tassiulas, L.~Jiang, P.~Basu, D.~Englund, and
  S.~Guha, ``Routing entanglement in the quantum internet,'' \emph{npj Quantum
  Information}, vol.~5, no.~1, pp. 1--9, Mar. 2019.

\bibitem{schoute2016shortcuts}
E.~Schoute, L.~Mancinska, T.~Islam, I.~Kerenidis, and S.~Wehner, ``Shortcuts to
  quantum network routing,'' \emph{arXiv preprint arXiv:1610.05238}, 2016.

\bibitem{quantum_queuing_delay}
W.~Dai, T.~Peng, and M.~Z. Win, ``Quantum queuing delay,'' \emph{IEEE Journal
  on Selected Areas in Communications}, vol.~38, no.~3, pp. 605--618, Feb.
  2020.

\bibitem{rezai2021quantum}
M.~Rezai and J.~A. Salehi, ``Quantum \text{CDMA} communication systems,''
  \emph{IEEE Transactions on Information Theory}, vol.~67, no.~8, pp.
  5526--5547, June 2021.

\bibitem{chen2019quantum}
X.-B. Chen, Y.-L. Wang, G.~Xu, and Y.-X. Yang, ``Quantum network communication
  with a novel discrete-time quantum walk,'' \emph{IEEE Access}, vol.~7, pp.
  13\,634--13\,642, Jan. 2019.

\bibitem{li2021building}
Z.~Li, K.~Xue, J.~Li, N.~Yu, J.~Liu, D.~S. Wei, Q.~Sun, and J.~Lu, ``Building a
  large-scale and wide-area quantum \text{Internet} based on an
  \text{OSI}-alike model,'' \emph{China Communications}, vol.~18, no.~10, pp.
  1--14, Nov. 2021.

\bibitem{nielsen_book}
M.~A. Nielsen and I.~L. Chuang, \emph{Quantum Computation and Quantum
  Information: 10th Anniversary Edition}.\hskip 1em plus 0.5em minus
  0.4em\relax Cambridge University Press, Dec. 2010.

\bibitem{chehimi2021entanglement}
M.~Chehimi and W.~Saad, ``Entanglement rate optimization in heterogeneous
  quantum communication networks,'' in \emph{Proceedings of the 17th IEEE
  International Symposium on Wireless Communication Systems (ISWCS)}, Sept.
  2021, pp. 1--6.

\bibitem{chehimi2022semantic}
M.~Chehimi, C.~Chaccour, and W.~Saad, ``Quantum semantic communications: An
  unexplored avenue for contextual networking,'' \emph{arXiv preprint
  arXiv:2205.02422}, 2022.

\bibitem{chehimi2021quantum}
\text{M. Chehimi and W. Saad}, ``Quantum federated learning with quantum
  data,'' in \emph{Proceedings of the 47th IEEE International Conference on
  Acoustics, Speech and Signal Processing (ICASSP)}, Singapore, May 2022, pp.
  1--5.

\end{thebibliography}


\end{document}